\documentclass[article]{spie}  

\usepackage{amsmath,amsfonts,amssymb}
\usepackage{graphicx}
\usepackage[colorlinks=true, allcolors=blue]{hyperref}
\usepackage{caption}
\usepackage{subcaption}
\usepackage{array, caption, floatrow, tabularx, makecell, booktabs}%
\usepackage{multicol, multirow}
\usepackage[style=ieee]{biblatex}
\bibliography{references} 
\newfloatcommand{capbtabbox}{table}[][\FBwidth]

\title{Increasing science yield with a twilight observing program with the SCALES instrument at Keck}

\author[a]{Isabel J. Kain}
\author[b]{Carlos Alvarez}
\author[b]{John M. O'Meara}
\author[b]{Marc Kassis}
\author[b]{Peter Wizinowich}
\author[b]{Antonin Bouchez}
\author[b]{Jim Lyke}
\author[b]{Randy Campbell}
\author[b]{Avinash Surendran}
\author[c]{Imke de Pater}
\author[c]{Katherine de Kleer}
\author[d]{Ned Molter}
\author[ ]{Erin Redwing}
\author[b]{Rosalie McGurk}
\author[a]{Steph Sallum}
\author[a]{Andrew Skemer}
\affil[a]{UC Santa Cruz}
\affil[b]{W. M. Keck Observatory}
\affil[c]{UC Berkeley}
\affil[d]{Space Telescope Science Institute}

\authorinfo{Corresponding author: Isabel J. Kain (ijkain@ucsc.edu)}

\begin{document} 
\maketitle

\begin{abstract}

While astronomical twilight closes the observing window for optical astronomers, the infrared sky remains dark even through sunrise, allowing IR astronomers to observe through twilight. The Slicer Combined with an Array of Lenslets for Exoplanet Spectroscopy (SCALES) instrument is a 2-5 micron coronagraphic integral field spectrograph scheduled to arrive at Keck in early 2026. SCALES has the potential to execute exciting science and support the astronomical community and upcoming NASA missions through a dedicated cadenced twilight observing program. We estimate that the current twilight observing program on Keck conducts 18$\pm$1 hours per year of science observations; a facilitized twilight observing program that is prioritized by the observatory could yield 151$\pm$2 hours of science time per year. This work presents the scientific motivation and high-level feasibility of two primary SCALES twilight science cases, monitoring of Solar System objects and a high-contrast imaging search for exoplanets around bright nearby stars, taking lessons from the existing NIRC2 and OSIRIS Twilight Zone program and considering increases in program scope. We also consider technical and operational challenges to overcome before the SCALES instrument begins its twilight observing program.

\end{abstract}

\keywords{Instrumentation, high-contrast, exoplanets, Solar System, infrared, integral field spectroscopy, Keck Observatory}

\section{Introduction}
\label{sec:intro}

\subsection{Motivation for a dedicated twilight observing mode at Keck and beyond}

The stable brightness of the infrared sky even as the sun approaches the horizon presents an opportunity for observatories to increase on-sky time and develop specialized observing modes that yield unique scientific results. The brightness of the sky above Maunakea in R-band changes by $\sim$6-8 magnitudes/arcsec$^2$ between dark time and sunrise (depending on moon phase), while M-band magnitudes remain constant to within $\lesssim$1 mag/arcsec$^2$ regardless of the Sun's position relative to the horizon \cite{aadamson_sites_2020}. This creates a window of opportunity, 40 mins in the evening and 40 mins in the morning, where infrared instruments can keep collecting photons where other instruments can’t. This presents not only additional time that can be appropriated for science, but a unique observing mode with relatively short exposures sampled at a $\sim$nightly cadence – which is perfect for characterizing objects that vary in time, or conducting efficient snapshot surveys of a large set of targets.

Some optical/infrared observatories even do daytime observing, including IRTF on Maunakea. This unique observing mode allows for an additional $\sim$6 hours of science time per each daytime observation, expanding the facility's available observing time by about 15\% \cite{john_rayner_infrared_2023}. Additionally, the best seeing typically occurs during moments of thermal equilibrium between air and ground, which tend to occur an hour before sunset and an hour after sunrise respectively \cite{bradley_characterization_2006}. While the development of a daytime observing mode for a 10-meter telescope presents additional operational and technological challenges that are outside the scope of this paper, this expansion of observing capabilities has been studied extensively, including for Keck Observatory (e.g. \cite{peter_wizinowich_lgs_2020, peter_wizinowich_keck_2020, dungee_feasibility_2019}).

In the age of 30-meter telescopes, 8-10-meter class facilities should invest in developing innovative observing strategies to maintain scientific productivity. In the age of extremely large-aperture telescopes such as E-ELT, existing facilities remain crucial assets of the scientific community, and innovative observing modes cement this. Clever observing schemes that maximize science output are necessary for the operations of modern observatories.

\subsection{Lessons learned from the existing Keck twilight program}
\label{subsec:tz}

Keck has executed a highly successful twilight observing program, The Twilight Zone \cite{alvarez_keck_2018}, which has monitored Solar System targets since the program's inception in 2018A (with test observations in 2017A and B). Twilight observations are executed with the OSIRIS instrument on Keck I or the NIRC2 instrument on Keck II. As of September 2025, this program has produced more than 200 observations, amounting to more than 100 hours of twilight operations. Six peer-reviewed papers have been produced thus far using results from this observing program (\cite{de_kleer_ios_2019}, \cite{molter_analysis_2019}, \cite{chavez_evolution_2023}, \cite{chavez_drift_2023}, \cite{nixon_atmosphere_2025}, \cite{i_de_pater_first_2025}). A second twilight program specifically for KPF has also operated at Keck – for the purposes of this paper, these programs are conflated in our discussion and analysis.

The Twilight Zone executes twilight observations opportunistically, when scheduled observers yield their time before 12$^{\circ}$ morning twilight, or the scheduled instrument experiences a failure that can’t be recovered before the end of the night. In these cases, and when weather conditions allow for continued science operation and either OSIRIS or NIRC2 are available, the Observing Assistant (OA, who operates the telescope and AO system on behalf of the observer) will switch to a twilight instrument, acquire the highest-priority Solar System target as determined by the scientific program, close the AO loop on the target itself, and execute an observing script that automatically configures the twilight instrument and takes a sequence of exposures.

As we think towards developing a dedicated twilight observing program for the SCALES instrument, and recommend Keck Observatory make programmatic changes to take advantage of twilight time, there are a number of lessons and best practices that can be gleaned from the successes of The Twilight Zone, largely due to the quality of observing tools and depth of collaboration with Keck OAs. 

Execution of twilight observations is made as frictionless as possible with a range of well-developed observing tools, including thorough step-by-step instructions for using each instrument for twilight observations, automated observing scripts, and easy-to-use observation planning tools including auto-generated observing charts for common Solar System targets, frequently updated priority lists for target and filter combinations, and tables showing how many exposures OAs can expect to capture given the amount of twilight time available. These tools, and the broader success of the Twilight Zone, can be attributed to a large investment of expert labor, including 5 distinct Keck Visiting Scholars, and close collaboration with the OAs who are responsible for executing the program.

However, the existing twilight program reveals opportunities for operational improvements as well. The two primary growth points are: (i) trading an opportunistic execution mode for programmatically prioritizing twilight observations, and (ii) balancing science programs to effectively utilize available twilight time year-round. 

A strength of the Twilight Zone program is that it operates opportunistically, conducting science operations during time that is \lq\lq thrown away\rq\rq  by scheduled observers. However, this leaves a large amount of science-capable twilight time underutilized, as we show in Section \ref{sec:time}. Keck Observatory could transition future twilight programs from opportunistic to classically scheduled, utilizing morning twilight time for science observations on nights that optical instruments are scheduled and deprioritizing non-science operations (e.g. instrument calibrations). 

A second growth point of a future twilight program is expanding beyond a Solar System-only science program. Taking 2026 as an example, Solar System targets are only observable for $\sim$half the year (see Figure \ref{fig:ss-targ-avail} and Section \ref{sec:science} for further discussion), which underutilizes this unique observing mode. Future twilight observing programs should consider science programs that take full advantage of available twilight time, as we explore in Section \ref{subsec:exoplanet}. 

As we seek to build a successful twilight observing program for the incoming SCALES instrument (see subsection \ref{subsec:SCALES}), the SCALES instrument and science team aim to:
\begin{itemize}
    \item Optimize instrument configuration and observing sequences to make efficient use of time-constrained twilight observations
    \item Develop frictionless set-and-forget user tools to streamline acquisition of twilight observations 
    \item Work closely with OAs to tailor observing tools to their needs
\end{itemize}

However, much of the success of a twilight observing program lies in facility prioritization (see Section \ref{sec:time} for a more granular discussion).

\subsection{Twilight program with SCALES}
\label{subsec:SCALES}

The Slicer Combined with an Array of Lenslets for Exoplanet Spectroscopy (SCALES) instrument \cite{skemer_design_2022, kupke_scales_2022, banyal_design_2022, stelter_scales_2024, surya_performance_2024} is an AO-fed high-contrast imaging instrument in development for the Keck II telescope, with expected delivery in January 2026. SCALES is comprised of a 1-5$\mu$m 12.3\rq\rq x 12.3\rq\rq imaging channel designed to replicate NIRC2’s capabilities, and a 2-5$\mu$m IFU, offering new capabilities compared to existing Keck instrumentation. SCALES’ sensitivity extends to longer wavelengths than existing high-contrast imaging IFUs (GPI, CHARIS, SPHERE) and is thus sensitive to older, colder exoplanets than have previously been directly imaged from the ground. While SCALES is optimized for exoplanet detection, it will enable a broad range of new science at Keck, including protoplanet characterization, protoplanetary disk mapping, Solar System object monitoring, and characterization of supernova remnants, active galactic nuclei, and more \cite{sallum_slicer_2023}.

\begin{figure}
     \centering
     \begin{subfigure}[b]{0.49\textwidth}
         \centering
         \includegraphics[width=\textwidth]{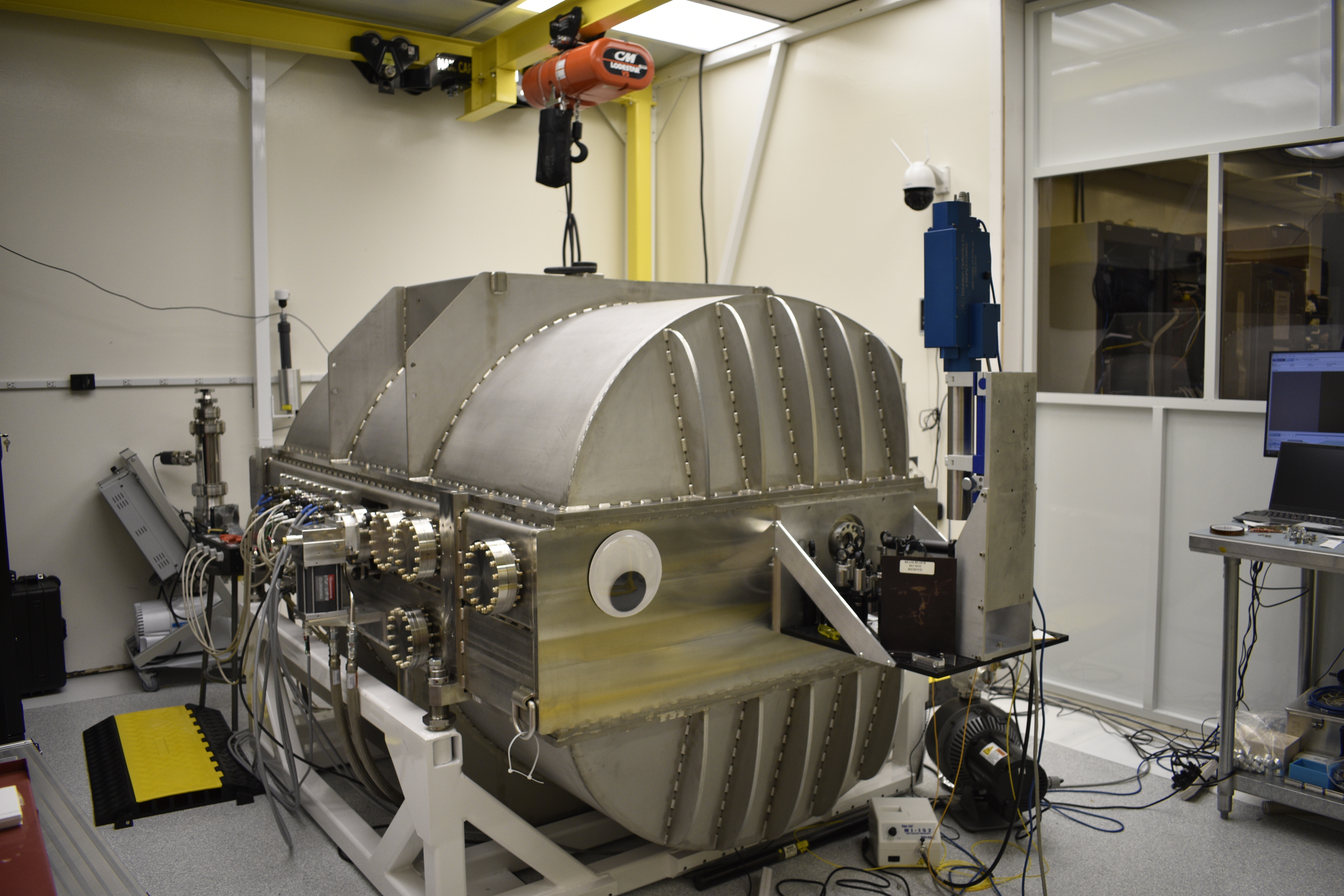}
     \end{subfigure}
     \hfill
     \begin{subfigure}[b]{0.49\textwidth}
         \centering
         \includegraphics[width=\textwidth]{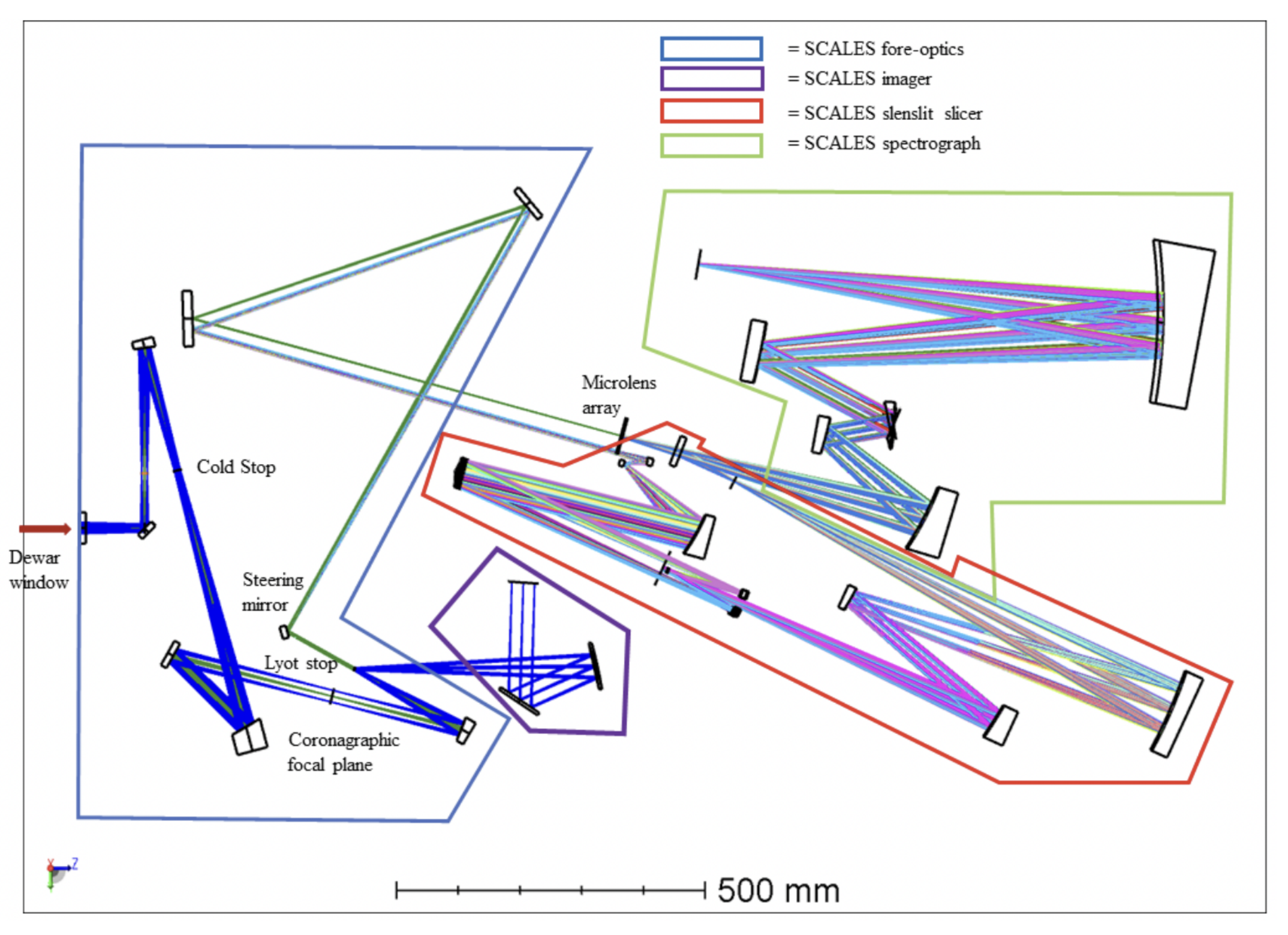}
     \end{subfigure}
        \vspace{2mm}
        \caption{SCALES instrument, Left: assembled cryostat going through a test cryocycle in the UCSC cleanroom. Right: SCALES optical design, with each optical subsystem outlined. SCALES is fed by the Keck II AO system (left side of diagram); the foreoptics (outlined in blue) relay the beam to the IFU or the imaging channel (outlined in purple). For the IFU, the tip-tilt steering mirror relays the beam directly onto the lenslet array (for the low-resolution IFU) or into the slenslit slicer (outlined in red); the beam then passes through the spectrograph (outlined in green) and is dispersed onto the detector. }
        \label{fig:scales}
\end{figure}

The IFU uses a combined slicer and lenslet array (“slenslit”) design described in \cite{stelter_colors_2021}, \cite{stelter_weighing_2022}, \cite{stelter_scales_2024}. The incoming beam from the Keck II AO system is relayed by the foreoptics onto a microlens array (Figure \ref{fig:scales}). The beam is focused into a grid of micro-pupils, which are passed through a pinhole grid to minimize cross-talk between adjacent lenslets. For the low-resolution mode, this 108x108 micro-pupil array is relayed directly through the spectrograph optics. For the medium-resolution mode, the slicer then reformats the 17x18 array of micro-pupils into a “pseudoslit,” where the 18 columns are rearranged into 3 vertical columns with smaller sub-groupings of the micro-pupils within each column, which then passes into the spectrograph to be dispersed. This reformatting is done to maximize the effective field of view of the IFU and prevent the dispersed micro-pupils (“spaxels” or spectral pixels) from overlapping on the detector (for a simulated SCALES datacube, see Figure \ref{fig:io-cube}).

Observers using the IFU can select between low (R $\sim$ 35-250, 2.2\rq\rq x2.2\rq\rq FOV) and medium spectral resolution (R $\sim$ 3500-7000, 0.36\rq\rq x0.34\rq\rq FOV) modes. A high-precision piezoelectric tip-tilt stage, located in the foreoptics subassembly, steers the beam between spectral resolution modes (Figure \ref{fig:scales}). In addition to toggling between resolution modes of the IFU, the tip-tilt stage can also be used for field stabilization or for taking mosaiced images of extended sources. 

As a future instrument upgrade of particular interest to twilight observing, a dichroic could be installed in the Lyot wheel (see Figure \ref{fig:scales}) to split light between the imaging channel and IFU, which would allow for simultaneous target acquisition in J- and H-band with the imager, and science integration in K- L- and M-band with the IFU. 

Upon arrival at Keck Observatory in early 2026, commissioning activities will include verifying SCALES performance in twilight observing mode and developing user tools to enable seamless SCALES twilight operations. All three SCALES instrument modes – the imaging channel, low-resolution IFU, and medium-resolution IFU – will be available for twilight observing, though this observing mode is sufficiently time-limited that observers should not plan on switching between multiple instrument channels. While instrument configuration overheads have been minimized through the design and integration process, this overhead is smaller than the time required to slew the telescope to the desired target and close AO loops and is therefore not a significantly limiting factor. Automated instrument configuration scripts can be set to run while the OA is slewing and closing loops to efficiently use twilight time for science. 

\section{On-sky time reclaimed by a facilitized Keck twilight observing program}
\label{sec:time}

To predict the time yield of the in-development SCALES twilight observing program, we examine two operational frameworks: the existing Twilight Zone program, which operates opportunistically when observers yield their time; and a dedicated twilight observing program that is programmatically prioritized on a wider range of scheduled nights. How much additional science time does the current twilight program yield, and how much additional science time could Keck Observatory reclaim with more substantive operational prioritization of this observing mode?

\subsection{Keck metrics analysis}
\label{subsec:metrics}

We analyze Keck metrics to benchmark the performance of the current Twilight Zone NIRC2 and OSIRIS program (\ref{subsec:tz-stats}), and better characterize the constraints on current twilight operations: early closures (\ref{subsec:es}), instrument scheduling (\ref{subsec:sched}), and weather (\ref{subsec:weather}). These estimates inform a final calculation of total available twilight time for the current and recommended operational frameworks (\ref{subsec:time-calc}).

The long-running Keck metrics effort was first piloted in 2001, with the goal of elucidating (and increasing) the fraction of observatory time spent collecting science photons. This program tracks how dark time is used, including telescope, instrument, and AO system activities (e.g. telescope focusing, slewing, instrument configuration, closing AO loops); instrument or facility faults; and time lost due to weather and engineering activities. This self-monitoring has succeeded in increasing Keck Observatory’s operational efficiency. 

Time spent in various states is scraped from instrument logs, meaning regular states – slewing, guiding, focusing, taking calibrations, taking science exposures, instrument fault, weather, etc. – are automatically collected and stored. Activity flags have a specified priority order: weather masks everything; engineering masks everything except weather; faults mask everything except weather and engineering; and so on. Currently, OAs manually log twilight observations through the night ticket system as “time lost,” even though this time is being appropriated for science purposes. Instrument schedules and weather statistics are also archived.

\subsubsection{Science time of current twilight program}
\label{subsec:tz-stats}

The Twilight Zone program (described in Section \ref{subsec:tz}) has shown to be highly scientifically productive: this program has executed 209 triggers between its inception and 2025A, amounting to at least 101 total hours of telescope time spent capturing twilight observations. 

While this program has been operational since 2018A, Keck Observatory operations were impacted by the Covid-19 pandemic, and so we limit the calculation of program performance statistics to 2021A-2025A. 

Within this time period, we find that twilight observations are triggered on 1.1$\pm$0.6\% of Keck I nights and 5.6$\pm$0.6\% of Keck II nights. Triggered twilight observations last for an average duration of 38$\pm$23 minutes per night. 

\begin{figure}
\begin{minipage}[c]{0.49\linewidth}
\includegraphics[width=\linewidth]{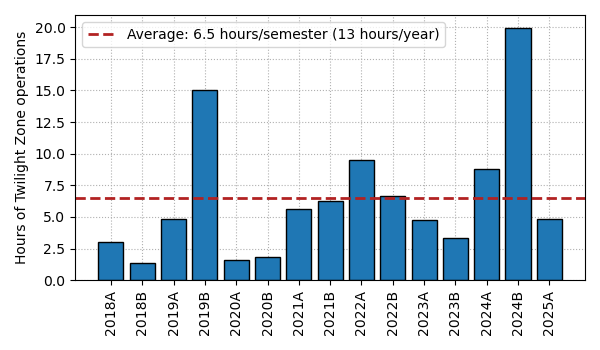}
\caption{Statistics for the current Twilight Zone program (opportunistic observations of Solar System objects with NIRC2 and OSIRIS), taken from 2018A through 2025A. Left: aggregate hours of operating time per semester. Right: distribution of time spent during individual twilight triggers, i.e. time spent per night on twilight observations.}
\label{fig:tz-time}
\end{minipage}
\hfill
\begin{minipage}[c]{0.49\linewidth}
\includegraphics[width=\linewidth]{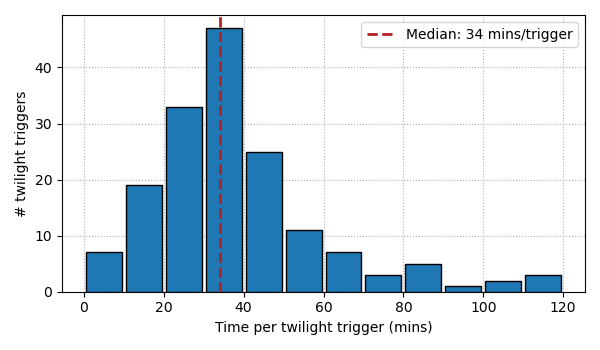}
\end{minipage}%
\end{figure}

A breakdown of the total time spent conducting Twilight Zone observations per semester, and statistics for time spent on twilight operations per night, can be found in Figure \ref{fig:tz-time}. However, these values should be taken as rough lower bounds on science time achieved with the Twilight Zone program. Many night tickets that log twilight observations list a \lq\lq Time lost\rq\rq  value of zero minutes even when observations are successful, with the rationale that this isn't observatory time that is wasted (as the Keck Metrics effort is built to track), but observatory time that has been reclaimed. Thus, the total number of twilight triggers is likely a more faithful estimate of program productivity.

\begin{figure}
\begin{floatrow}
\capbtabbox{
    \renewcommand{\arraystretch}{1.2} 
    \begin{tabular}{c c c}
    \toprule
    \toprule
      \textbf{Telescope} & \textbf{Instrument} & \textbf{Twilight Triggers}\\
      \cmidrule(lr){1-3}
      \multirow{7}{*}{Keck I} & HIRES & 5 \\ 
      & KPF & 2 \\ 
      & LRIS & 10 \\
      & MOSFIRE & 0 \\
      & OSIRIS & 11 \\
      & PCS/SSC & 5 \\
      \cmidrule(lr){3-3}
      &  & Total: 33 \\
      \cmidrule(lr){1-3}
      \multirow{7}{*}{Keck II} & DEIMOS & 74 \\
      & ESI & 0 \\
      & KCWI & 35 \\ 
      & NIRC2 & 30 \\
      & NIRES & 13 \\
      & NIRSPEC & 15 \\
      & PCS/SSC & 9 \\
     \cmidrule(lr){3-3}
      & & Total: 176 \\
    \bottomrule
    \bottomrule
    \end{tabular}
  }{\caption{Frequency of twilight triggers by second-half scheduled instrument between 2016B (though most twilight observations are after 2018A when the program officially began operations) and 2025A. The KCWI statistics include both KCWI and KCRM scheduling, and the NIRSPEC statistics include NIRSPEC, NIRSPAO, and KPIC scheduling.}
  \label{tab:trigger-stats}
  }
\ffigbox{
  \includegraphics[width=0.45\textwidth]{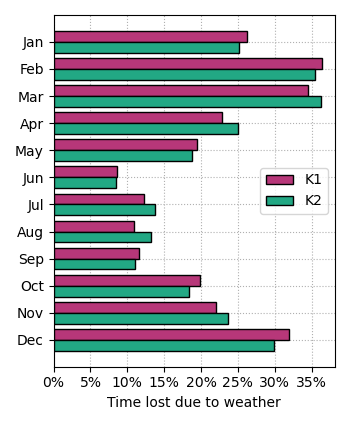}
  }{
  \caption{Average percent of dark time lost to weather-related closures between 2018A and 2025A.}
  \label{fig:weather}
  }
\end{floatrow}
\end{figure}

Table \ref{tab:trigger-stats} shows statistics for which second-half scheduled instruments coincide with triggered twilight observations. To demonstrate the information contained in this table, the first row indicates that between 2018A and 2025A, HIRES was the scheduled second-half instrument for 5 nights where twilight observations were executed.

There is a clear under-utilization of Keck I instruments for Twilight Zone operations; because of small-number statistics, it's difficult to draw conclusions about instrument useage for twilight operations. LRIS is a common second-half instrument during twilight triggers, likely because it is a visible-light instrument. OSIRIS is likely favored because of lower observational overhead for a twilight trigger (i.e. the OA does not need to switch to a different instrument, and can instead simply slew the telescope to a twilight target and take data). 

Keck II OAs triggers twilight observations much more frequently, with DEIMOS, KCWI, and NIRC2 being the most common second-half instruments when twilight observations are triggered. DEIMOS and KCWI, as visible-light instruments, are likelier to yield their twilight time because their science capabilities are constrained under a brightening twilight sky; NIRC2 is likely a common second-half instrument because an instrument switch is unnecessary, similar to OSIRIS on Keck I. Certain instruments are never utilized for twilight observations because an on-the-fly instrument switch is impossible. There may be a reduction in the frequency of KCWI observers yielding twilight time, given the commissioning of the Keck Cosmic Reionization Mapper (KCRM), the newly-commissioned red arm of KCWI which extends out to 1$\mu$m. 

It's clear that some infrared instruments such as NIRES and NIRSPEC, while nominally capable of observing into twilight, are yielding their twilight time for Twilight Zone operations, indicating an opportunity for instruments with specifically designed twilight capabilities and science programs to step in and utilize this time. 

\subsubsection{Early shutdowns}
\label{subsec:es}

Early shutdowns (ES) occur when observers or observatory staff choose to end the night before 12 degree twilight (this does not include instrument or facility failures or weather-related closures). These events potentially represent time that the observatory could have spent executing twilight observations; here we briefly explore what proportion of these early shutdown events could be reclaimed by an opportunistic twilight observing program. 

Early shutdown statistics, and our estimate of productive twilight science time that is recoverable from these early shutdowns, are summarized in Table \ref{tab:early-shut}. 220 early shutdowns have been logged between 2018A and 2025A, with an average rate of 29$\pm$5 per year including both telescopes. Assuming that a twilight observation can be accomplished with a minimum of 15 minutes (a benchmark number that is not rigorously selected), one-third of these early shutdowns could have instead been appropriated for twilight observations, amounting to an average of 5.2$\pm$1.0 hours per year of reclaimed science time between both telescopes. This value is estimated from: the average rate of early shutdowns; the proportion of early shutdowns that are twilight-capable (i.e. with a duration of 15 minutes or longer); and the average duration of twilight-capable early shutdowns. 

\smallskip 
\renewcommand{\arraystretch}{1.2} 
\begin{table}[H]
\begin{tabular*}{\columnwidth}{@{\extracolsep{\fill}}l ccc}
\toprule
\toprule
     & \textbf{Both} & \textbf{K1} & \textbf{K2} \\
  \cmidrule(lr){2-4}

  Average early shutdowns per year & 29$\pm$5 & 12$\pm$3  & 18$\pm$4 \\
  Average time lost per early shutdown (mins) & 16$\pm^{28}_{16}$ & 16$\pm^{30}_{16}$    & 15$\pm^{27}_{15}$       \\
  Time lost to early shutdowns (hours/year) & 7.6$\pm$0.6 & 3.2$\pm$0.5   & 4.4$\pm$0.6      \\ 
  
  \cmidrule(lr){1-4}

\textbf{\textit{Time recoverable for twilight operations (hours/year)}} & \textbf{\textit{5.2$\pm$1.0}} & \textbf{\textit{2.3$\pm$1.0}}  & \textbf{\textit{2.9$\pm$0.9}}      \\
\bottomrule
\bottomrule
\end{tabular*}
{\caption{Early shutdown statistics for the Keck I and Keck II telescopes, taken from 2018A through 2025A.}
\label{tab:early-shut}
}
\end{table}

\subsubsection{Instrument scheduling}
\label{subsec:sched}

Which instrument is scheduled for the second half-night affects the ability of any twilight observing program to execute. While the scheduled second-half instrument does not need to be capable of twilight observations, it must be possible for the OA to switch to a twilight-capable instrument without prohibitive overhead. ESI sits at Cassegrain focus, and cannot be swapped with a twilight-capable instrument because an instrument swap requires hours of work from the summit day crew. Conversely, NIRSPAO has yielded a large number of twilight observations to date because the Keck II beam can be redirected from NIRSPAO to NIRC2 by rotating a single fold mirror in the Keck II AO system.

Another constraint on twilight observations is the ability of the scheduled second-half instrument to continue its nominal science program into twilight time. Some infrared instruments can and do continue collecting science photons well into 12-degree twilight, though as discussed in Section \ref{subsec:tz-stats}, many infrared instruments still yield their twilight time for use in the Twilight Zone program. Therefore, it is not required that scheduled second-half instruments must be incapable of doing science during twilight (e.g. optical instruments) for execution of a twilight observing program to be possible during that night. 

\begin{figure}[ht]
     \centering
     \begin{subfigure}[b]{0.49\textwidth}
         \centering
         \includegraphics[width=\textwidth]{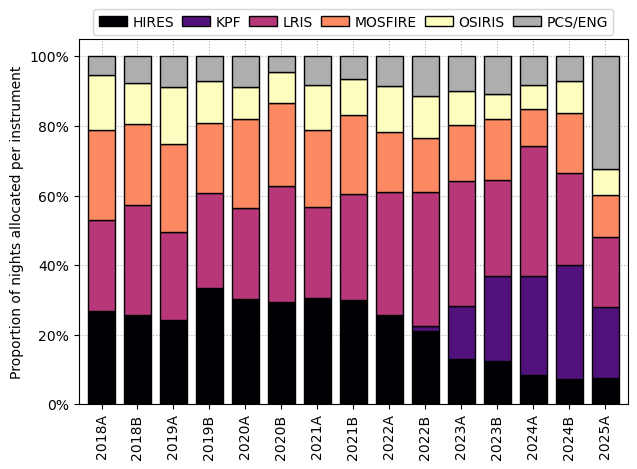}
     \end{subfigure}
     \hfill
     \begin{subfigure}[b]{0.49\textwidth}
         \centering
         \includegraphics[width=\textwidth]{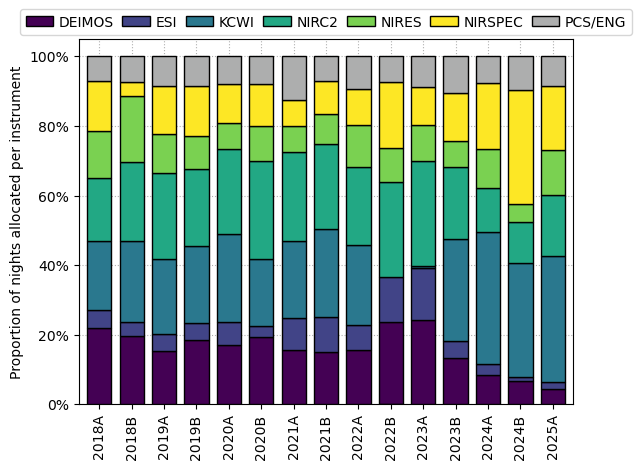}
     \end{subfigure}
        \vspace{2mm}
        \caption{Instrument scheduling trends between 2018A and 2025A (left: Keck I, right: Keck II). }
        \label{fig:scheduling}
\end{figure}

We analyzed Keck instrument scheduling data between 2018A and 2025A to determine the typical fraction of nights every semester that instruments are scheduled to be on-sky. The full-baseline instrument scheduling trends are shown in Figure \ref{fig:scheduling}; a summary of instrument scheduling statistics, taken over a more recent time period of 2023A through 2025A, is shown in Table \ref{tab:sched}. This shorter baseline is used to estimate typical instrument scheduling statistics because of fluctuations in instrument scheduling caused by new instrument arrivals and facility instrument upgrades. Figure \ref{fig:scheduling} shows that 2023A is the beginning of a relatively homogeneous period of instrument scheduling trends, since this is roughly when both KCRM and KPIC commissioning concluded.

\renewcommand{\arraystretch}{1.2} 
\begin{table}[h!]
  \begin{center}
    \caption{Summary of instrument scheduling statistics for the Keck I and Keck II telescopes, taken from 2023A through 2025A. Time period was chosen to account for facility instrument changes, e.g. KCRM and KPIC commissioning. Estimated percent is taken from instrument scheduling averages, and error bars are taken from the standard deviation of percent scheduled nights per semester. Scheduled PCS and ENG time is not relevant here and thus neglected.}
    \label{tab:sched}
    \begin{tabular}{c c c}
    \toprule
    \toprule
      \textbf{Telescope} & \textbf{Instrument} & \textbf{Percent nights scheduled per semester} \\
      \cmidrule(lr){1-3}
      \multirow{5}{*}{Keck I} & HIRES & 9.8$\pm$2.8\%\\ 
      & KPF & 24.3$\pm$6.8\% \\ 
      & LRIS & 29.5$\pm$7.1\% \\
      & MOSFIRE & 14.6$\pm$3.2\% \\
      & OSIRIS & 8.1$\pm$1.2\% \\
      
      \cmidrule(lr){1-3}
      
      \multirow{6}{*}{Keck II} & DEIMOS & 11.4$\pm$8.0\% \\
      & ESI & 5.3$^{5.6}_{5.3}$\% \\
      & KCWI/KCRM & 27.3$\pm$15.3\% \\ 
      & NIRC2 & 18.6$\pm$7.5\% \\
      & NIRES & 9.5$\pm$3.2\% \\
      & NIRSPEC/NIRSPAO/KPIC & 18.9$\pm$8.4\% \\
    \bottomrule
    \bottomrule
    \end{tabular}
  \end{center}
\end{table}

The twilight statistics presented in Section \ref{subsec:tz-stats}, as well as conversations with Keck OAs, suggest that HIRES and LRIS on Keck I and DEIMOS and KCWI on Keck II are most amenable to opportunistic twilight observations when the scheduled observers finish early. We note that these frequently-used twilight instruments are respectively scheduled for 38.7$\pm$17.3\% of Keck I nights and 39.3$\pm$7.6\% of Keck II nights per semester. 

\subsubsection{Weather}
\label{subsec:weather}

As one of the factors that observatory staff cannot control, average time lost to weather is the lower bound on observatory efficiency. Figure \ref{fig:weather} shows the monthly average percentage of telescope time lost due to weather from 2018A through 2025A. Lost time is reported in duration of weather-related closures. To obtain percentages of time lost, we calculate the total dark time per month (all time where the sun is greater than 12$^{\circ}$ below the horizon, which includes both true nighttime and astronomical twilight) using the JPL Horizons tool \cite{giorgini_status_2015}. This relies on the assumption that weather-related closures are almost always only logged as time lost during dark time.

Averaged across this eight-year period and both telescopes, weather accounts for a loss of 21.7$\pm$17.4\% of observing time, with no significant difference in weather loss between A and B semesters. 


\subsection{Estimating productivity of current and future twilight observing programs}
\label{subsec:time-calc}

We use our analysis of historical trends in Twilight Zone operations, instrument scheduling, and weather-related time loss to calculate current and projected operational time, in hours per year, of three twilight program frameworks (with our findings summarized in Table \ref{tab:time-results}):
\begin{enumerate}
    \item The current Twilight Zone observing program (opportunistic observing with NIRC2 and OSIRIS) with no proposed changes 
    \item The current Twilight Zone observing program (opportunistic observing with NIRC2 and OSIRIS), assuming that all twilight-capable early shutdowns are appropriated for twilight operations
    \item A future, facilitized twilight observing program that is scheduled for operation during every available morning
\end{enumerate}

As enumerated in Section \ref{subsec:tz}, we estimate that the current Twilight Zone program averages 18$\pm$1 hours per year of science time. If we assume a continuation of the opportunistic operation framework of the current Twilight Zone program, but push to use every minute of science-capable twilight time for Twilight Zone operations (i.e. re-appropriating all sufficiently long early shutdowns that are not related to weather, instrument failures, or facility failures), then the time yield of this observing mode could be increased to 23$\pm$1 hours per year of twilight science. However, without making substantive operational changes, any future twilight observing program cannot make more than marginal gains of a few hours per year in observing time.

To realistically estimate the time yield of a facilitized twilight observing program, we start with the most generously circumscribed window of \lq\lq available\rq\rq  twilight time, and apply progressively tighter and statistically-motivated constraints. 

Throughout the year, the period between 12$^{\circ}$ and 3$^{\circ}$ twilight (at which point the Earth's atmosphere begins to refract sunlight over the horizon) varies from 38 to 44 minutes in duration, with an annual average of $\sim$40 minutes. This is the ceiling of twilight time recoverable by any twilight observing program.

We assume that only morning twilight will be utilized for any twilight observing program. While evening twilight presents all the opportunities and advantages outlined above, evening twilight time is generally occupied by facility activities that are difficult to relocate (e.g. telescope focusing). Future work could be useful in clearing evening twilight of all non-science activities to maximize twilight science yield.

The only constraint that is fully outside of the observatory's control is weather. As we found in Section \ref{subsec:weather}, 21.7$\pm$17.4\% of Keck time is lost due to weather. 

To approximate a scenario where twilight observing is prioritized for every clear night when instrument scheduling constraints allow, we assume all nights where the most common second-half instruments for Twilight Zone observations (HIRES and LRIS on Keck I, DEIMOS and KCWI on Keck II) are utilized for twilight observing. This is an extremely rough estimation, since we expect 1) realistically, less than 100\% of observing nights on any instrument will be allocated to even an appropriately prioritized twilight observing program, and 2) an appropriately prioritized twilight observing program will also execute when other instruments are scheduled for the second half. We expect that this over- and underestimation, respectively, roughly balance each other. For these reasons, we stress that the calculation presented in this section is intended as a guide, and not a high-fidelity prediction. This estimation yields 39.3$\pm$7.6\% of Keck I time, and 38.7$\pm$17.3\% of Keck II time allocated for a facilitized twilight observing program.


Assuming an average twilight window of 40 minutes, and factoring in the above constraints – morning twilight only, accounting for average weather, and assuming allocation of 39.3$\pm$7.6\% of Keck I time and 38.7$\pm$17.3\% of Keck II time – a facilitized twilight observing program could yield an estimated 151$\pm$2 hours per year of science time. This is roughly the equivalent of 15 Keck nights. 

\renewcommand{\arraystretch}{1.2} 
\begin{table}[H]
  \begin{center}
    \caption{Estimated productivity of three twilight observing program operational frameworks: (1) the current Twilight Zone program, which conducts opportunistic Solar System monitoring with OSIRIS and NIRC2; (2) a slightly more efficient version of the current Twilight Zone program, where all early shutdowns (where observers voluntarily end the night early unrelated to weather, instrument failures, or facility failures) are utilized instead for twilight observing; and (3) a facilitized twilight observing program with substantive operational modifications.}
    \label{tab:time-results}
    \begin{tabular}{l l c c}
    \toprule
    \toprule

    1. & The Twilight Zone & 18$\pm$1 hours/year & $\sim$2 Keck nights \\
    2. & The Twilight Zone, re-appropriating some early shutdowns & 23$\pm$1 hours/year & $\sim$2 Keck nights \\
    \textbf{3.} & \textbf{Facilitized twilight observing program} & \textbf{151$\pm$2 hours/year} & \textbf{$\sim$15 Keck nights} \\
    

    \bottomrule
    \bottomrule
    \end{tabular}
  \end{center}
\end{table}

\section{Science programs}
\label{sec:science}

Twilight observing presents a unique science mode. With time-limited observations that can occur at a cadence of (ideally) every morning, twilight observing lends itself well to monitoring of time-domain phenomena, or large-scale surveys with relatively short observations.

\begin{figure}
     \centering
     \begin{subfigure}[b]{1\textwidth}
         \centering
         \includegraphics[width=\textwidth]{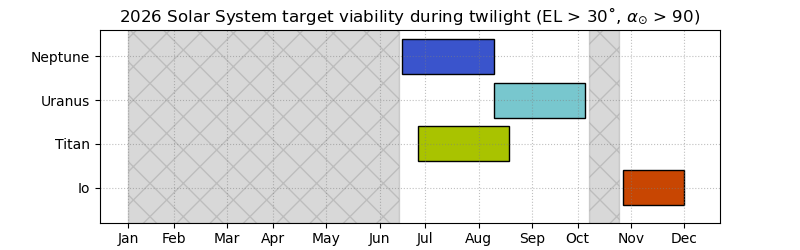}
    \end{subfigure}
        \caption{Availability of four representative Solar System targets – Io, Titan, Neptune, and Uranus – throughout 2026. A target is considered available if it is $\geq$30$^{\circ}$ above the horizon and $\geq$90$^{\circ}$ away from the rising sun, calculated at sunrise (3$^{\circ}$ twilight, at which point Earth's atmosphere begins refracting sunlight over the horizon).}
        \label{fig:ss-targ-avail}
\end{figure}

The current Twilight Zone program has demonstrated that Solar System monitoring is an excellent science case. We explore SCALES' capabilities for a twilight Solar System monitoring program in the following section (\ref{subsec:ss}). However, as mentioned in Section \ref{subsec:tz}, a set of science programs that exclusively focuses on Solar System targets does not take full advantage of twilight time. Figure \ref{fig:ss-targ-avail} shows the availability of four representative Solar System targets – Io, Titan, Neptune, and Uranus – throughout 2026. A target is considered available if it is $\geq$30$^{\circ}$ above the horizon and $\geq$90$^{\circ}$ away from the rising sun, calculated at sunrise (3$^{\circ}$ twilight), which minimizes the risk of sunlight glancing off the Keck primary mirror. Because of these constraints, no Solar System targets are available to observe for more than half the year (195 nights in 2026 where zero Solar System targets are accessible). To take full advantage of twilight time, future twilight observing programs – including the program in development for SCALES – should consider combining complementary science programs that fully utilize twilight time. There are many science programs that fit this specification, but for the purposes of this paper we explore an exoplanet direct imaging survey (Section \ref{subsec:exoplanet}).

\subsection{Solar System monitoring}
\label{subsec:ss}

Our nearest planetary neighbors present a testbed for understanding time-domain planetary-scale processes, which answers questions both about the Solar System bodies themselves and their more distant exoplanet cousins. In the following section, we explore high-level SCALES twilight capabilities for observing volcanism on Io, weather on Titan and the ice giants, and transient water plumes on Enceladus and Europa. This is by no means intended to be an exhaustive list of Solar System science cases for SCALES twilight observations, but an exploration of SCALES' applicability to some of the priority science questions e.g. from the 2023-2032 Planetary Decadal Survey \cite{committee_on_the_planetary_science_and_astrobiology_decadal_survey_origins_2023}.

Twilight monitoring of Solar System targets with SCALES also provides crucial support to NASA mission development. A number of NASA SMD missions have been identified as priorities for development and launch within the next few decades. Europa Clipper, on its way to characterize Jupiter's icy moons, launched in 2024. The Dragonfly rotorcraft mission, slated for launch in 2028 and arrival in 2032, will explore the dry equatorial regions of Titan. The 2023-2032 Planetary Decadal Survey \cite{committee_on_the_planetary_science_and_astrobiology_decadal_survey_origins_2023} has identified the Uranus Orbiter and Probe (UOP) mission concept as the highest priority flagship mission of the next decade, followed in priority by the Enceladus Orbilander. The Io Volcano Observer space mission has been selected for Phase A study. As these and other missions prepare for launch and science operations, insight into the time-variable properties of their destinations will be crucial for selecting probe sites, informing mission operations, and more.

Finally, we highlight a synergy between JWST and SCALES twilight observations of Solar System objects. NIRSpec on JWST and SCALES on Keck share a similar wavelength coverage ($\sim$1-5$\mu$m) and spectral resolution (NIRSpec R$\sim$100-2700, SCALES R$\sim$35-7000); however, the larger telescope aperture (10 meters compared to 6) grants SCALES superior spatial resolution. Many Solar System targets have been studied by JWST just within the first few observing cycles (e.g. \cite{melin_discovery_2025, villanueva_jwst_2023, nixon_atmosphere_2025}), with many emphasizing a JWST-Keck complement as a crucial strategy for groundbreaking Solar System science. Keck observing time is more available and flexible than JWST time; this is doubly the case for the twilight observing mode, which affords a high-cadence and low-overhead Keck observing contribution.

\subsubsection{Io's volcanoes}

Io is the only body in our Solar System where volcanism has been observed and is currently experiencing tectonic processes, granting astronomers a unique window through which to study solid body interiors. Its strong internal heating and activity make Io an important analogue to young terrestrial bodies – including the earliest years of Earth and its Moon – and tidally heated exoplanets. While some of Io’s volcanoes show persistent activity across decades, numerous transient eruptions appear and subside on timescales of days to minutes. Constraining Io's volcanism - including the spatial distribution, time variability, and temperature of eruptions, as well as their interaction with the atmosphere - is key to characterizing Io as a dynamic geophysical testbed. 

Long-baseline monitoring of Io's volcanic eruptions, most notably through the current Twilight Zone program with Keck, has provided key insights into Io's interior dynamics \cite{de_kleer_ios_2019}. However, many questions remain that can be addressed with continued monitoring, leveraging SCALES' expanded instrument capabilities compared to existing Keck instrumentation.

\begin{figure}[H]
     \centering
     \begin{subfigure}[b]{1\textwidth}
         \centering
         \includegraphics[width=\textwidth]{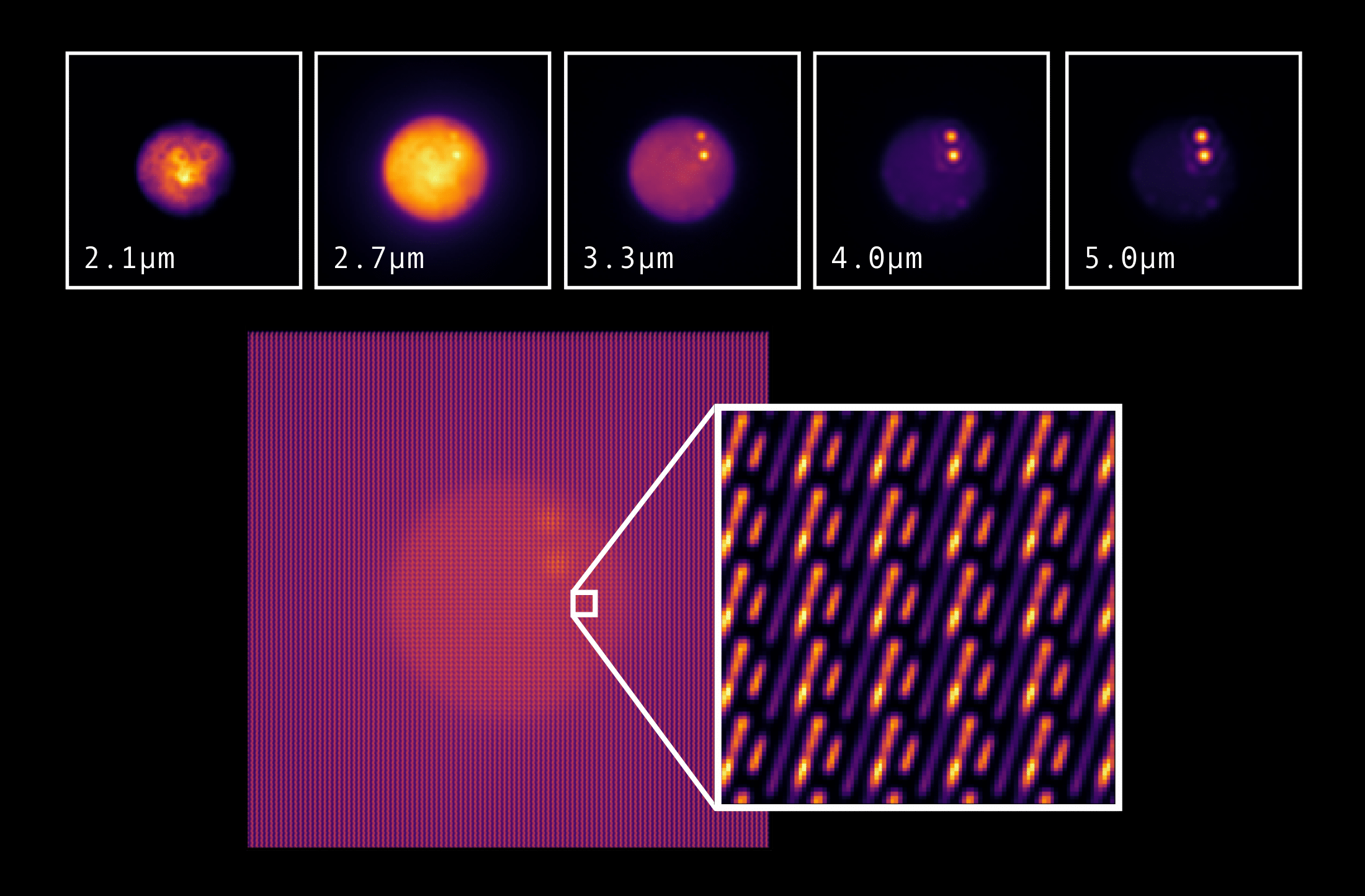}
    \end{subfigure}
        \caption{Simulated SCALES low-resolution IFU observations of Jupiter's moon Io. Simulations generated with \texttt{scalessim} with real Twilight Zone images taken with NIRC2, which are made available to the community \cite{noauthor_twilight_nodate}. Top: slices from simulated SCALES datacube of Io, with the wavelength of each slice annotated. Bottom: a raw SCALES low-resolution IFU data product, before being processed and assembled into a datacube. The cutout zooms in on the raw image, showing the \lq\lq spaxels\rq\rq  (spectral pixels) dispersed along the SCALES detector. Shorter wavelengths reveal surface features in reflected light, while longer wavelengths reveal thermal emission from Io's volcanoes.}
        \label{fig:io-cube}
\end{figure}

SCALES' 12.3\rq\rq x 12.3\rq\rq field-of-view imaging channel has spatial sampling of 0.006\rq\rq x0.006\rq\rq, and both IFU resolution modes are Nyquist sampled at 2$\mu$m with resolution elements of 0.02\rq\rq x0.02\rq\rq. This spatial sampling can resolve individual surface features as small as $\sim$150km at 2$\mu$m (with slightly coarser spatial resolution at longer wavelengths sensitive to the thermal emission of eruptions), enabling site-specific monitoring.

With wavelength coverage from 2-5$\mu$m, SCALES IFU modes will enable more robust constraints of volcano
temperatures than broadband imaging. SCALES can sufficiently sample the SED of Io's volcanoes as cool as $\sim$300$^{\circ}$K and constrain their temperature to within $\sim$30$^{\circ}$K \cite{sallum_slicer_2023}. The low-resolution IFU has a dedicated SED mode, spanning a full wavelength range of 2.0-5.0 $\mu$m with R$\sim$35, which will be able to characterize the full emission spectrum of every (sufficiently bright) volcano across Io's face with a single observation, which is an especially well-suited capability for time-limited twilight observations. Figure \ref{fig:io-cube} shows an simulated SCALES low-resolution IFU observation of Io in SED mode, generated with the \texttt{scalessim} observation simulation tool \cite{briesemeister_end--end_2020}.

\subsubsection{Weather on Titan}


Titan is the only body in our Solar System, other than Earth, known to sustain an active hydrologic cycle. Methane storms have been shown to strongly impact Titan's surface. Fluvial features were detected by the Huygens Probe even at dry mid-latitudes\cite{soderblom_topography_2007} that, until recently, were thought to be incapable of supporting convection, implying that these precipitation events are sufficiently periodic and intense to influence the surface features of these regions \cite{turtle_rapid_2011}. 

Previous monitoring programs have detected transient cloud formation events that resulted in global weather events, even in regions that had been dry and cloud-free for years \cite{schaller_storms_2009}. Characterization of the characteristics and tendencies of these storms over time has been pursued \cite{olim_methane_2025}, but precise knowledge of where and when these storms occur and evolve require long-baseline studies that span Titan's seasons. This time-domain characterization is crucial to informing development and operations of probes such as the Dragonfly mission.

SCALES will be able to spatially and spectrally resolve Titan's near-infrared weather features with twilight-compatible short exposure times \cite{bailey_near-ir_2011, baines_atmospheres_2005}. Various molecular features are distributed throughout SCALES wavelength coverage of 2-5$\mu$m, including highly reflective methane clouds as well as H$_2$, CH$_3$D, and CO , which are accessible with the spectral resolution of the low-resolution IFU. Different wavelengths probe different depths in the atmosphere, allowing IFU observations to not only detect the presence and composition of atmospheric features, but also simultaneously constrain their spatial location and extent and altitude. 

\subsubsection{Storms on Neptune and Uranus}

As nearly 6000 exoplanets have been discovered to date, planets between the sizes of the ice giants and Earth (dubbed mini-Neptunes or super-Earths) have emerged as the most common demographic. However, of the subset of the mini-Neptune population with both mass and radius constraints,  there is a broad range in density that spans the ice giants, indicating that both Uranus and Neptune are crucial cyphers for understanding this transition in density from hydrogen-dominated to rock-dominated worlds \cite{wakeford_exoplanet_2020}. However, both Uranus and Neptune are critically under-characterized members of the Solar System, with time-domain atmospheric and interior dynamics that have remain severely undersampled that stymie attempts to connect bulk compositions to more interesting conclusions of atmospheric dynamics and chemistry, interior structure, and dynamical history within the context of their formation in the Solar System. 

Beyond general relevance to both planetary and exoplanet science priorities, both Uranus and Neptune require long baseline monitoring to reveal their interior structure, atmospheric dynamics, and dynamical history. The only remote sensing measurements of Uranus were taken by Voyager 2 in 1986, which measured negligible heat flux from the deep interior \cite{pearl_albedo_1990}. This result has a range of implications. If storms and cloud-forming events are common and widespread, then heat is being efficiently convected from the deep interior to the upper atmosphere, and a low heat flux implies that Uranus experienced some early loss of internal heat. Conversely, if these storms are sporadic or regionally limited, then Uranus’ interior is not fully convective, suggesting that internal heat is being restricted to the deep interior where the Voyager probe would not be able to sense it \cite{wang_internal_2025}. 

Additionally, Uranus’ high 98$^{\circ}$ obliquity means each pole experiences 42 years of continuous sunlight, followed by 42 years of continuous darkness. This strong imbalance between internal and radiative heating means Uranus’ weather is primarily shaped by seasonal forcing. However, studying seasonal variations is challenging given its 84 year orbital period and $\sim$20 year seasonal durations. We are now half a Uranian year from the Voyager 2 measurements, meaning cadenced monitoring of Uranus’ weather will characterize a part of Uranus’ seasonal cycle that has not yet been probed by modern astronomical instruments. 

Neptune's atmosphere is highly active and chaotic, with highly-structured storm systems appearing and dispersing over timescales ranging from hours to decades. Despite nearly 30 years of high spatial resolution monitoring of Neptune by ground-based telescopes across nearly 3 solar cycles, Neptune has only been observed through $\sim$20\% of its orbit \cite{chavez_evolution_2023}. Continued monitoring in the infrared at high sampling cadences and long baselines is required to detect and characterize these storms as the form and evolve.

Thermal emission from Neptune itself is largely suppressed in the near-infrared. Methane clouds, however, are highly reflective in this bandpass – in K-band (2.2 $\mu$m), the contrast in reflectivity between methane clouds and regions free of tropospheric clouds reaches up to two orders of magnitude \cite{de_pater_neptunes_2014}. High-cadence near-infrared observations are useful to probe the structure, composition, and evolution of these storms \cite{hueso_monitoring_2020}, some of which have been discovered and characterized through Twilight Zone monitoring \cite{molter_analysis_2019}.

While SCALES' imaging mode can extend the successes of NIRC2 twilight monitoring of Uranus and Neptune, SCALES' IFU allows for simultaneous sampling of the spectral features and spatial distribution of weather and clouds across the planetary disk, enhancing the information content of twilight observations. SCALES’ R$\sim$200 K-band and R$\sim$250 CH4 low-resolution modes will provide the required wavelength coverage and spectral resolution to measure methane features at $\sim$2-2.25$\mu$m and $\sim$3-3.5$\mu$m on Uranus and Neptune. 

\begin{figure}
     \centering
     \begin{subfigure}[b]{0.49\textwidth}
         \centering
         \includegraphics[width=\textwidth]{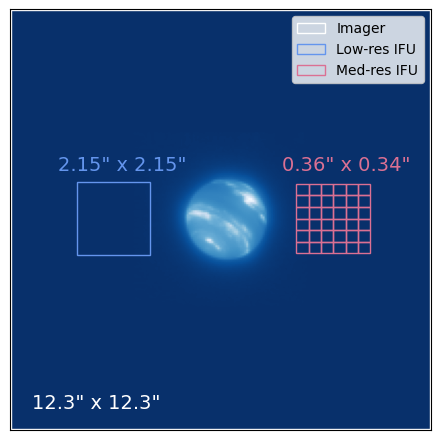}
     \end{subfigure}
     \hfill
     \begin{subfigure}[b]{0.49\textwidth}
         \centering
         \includegraphics[width=\textwidth]{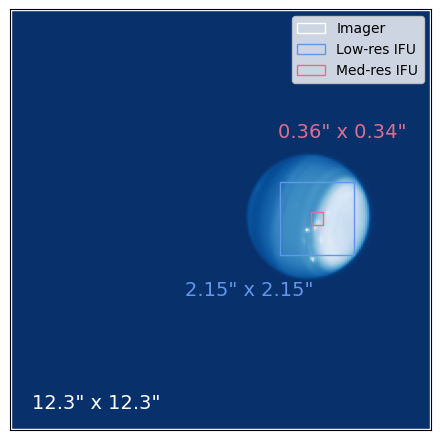}
     \end{subfigure}
        \caption{Simulationed SCALES observations of Neptune (left) and Uranus (right), incorporating real NIRC2 twilight observations which are made available to the community \cite{noauthor_twilight_nodate}. The fields of view of different SCALES channels (the imager, low-resolution IFU, and medium-resolution IFU) are overplotted for reference. The tiling of the medium-resolution mode FOV reflects the scanning capability of the internal tip-tilt stage, allowing SCALES to stitch together mosaic observations of an extended source. Simulations generated with \texttt{scalessim}.}
        \label{fig:uranus-neptune}
\end{figure}

Figure \ref{fig:uranus-neptune} shows \texttt{scalessim}-generated SCALES imaging channel observations of Uranus and Neptune. The field-of-view of each of the three SCALES instrument modes is annotated on each simulated observation. While SCALES' imaging mode can capture the entire planetary disk within its larger field-of-view, observers can choose the spectral advantage of either IFU mode without sacrificing effective field-of-view. SCALES has an internal tip-tilt stage that is primarily for steering the beam between the low- and medium-resolution IFU modes, but can also be used to steer the image on the detector. The tip-tilt stage could be used to mosaic individual pointings into a larger effective field-of-view. The tip-tilt stage is expected to have a dynamic pointing range of $\sim$3\rq\rq on-sky, meaning 2 low-res IFU pointings or $\sim$8 med-res IFU pointings could be stitched together along each axis to spatially sample the entire disk while maximizing spectral sampling. The instrument overhead for this observing configuration is expected to be no slower than taking consecutive exposures without using the tip-tilt stage to tile the observations, making this observing configuration a possibility for sampling extended objects even during time-limited twilight observations.


\subsubsection{Icy plumes from Enceladus and Europa}

Of the five confirmed ocean moons of our Solar System, two of them – Enceladus and Europa – have been observed to eject icy plumes from their subsurface oceans. These plumes imply geophysical activity, including cryovolcanism and tectonism, and offer a window into the composition of otherwise hidden interior seas. 

Recently, JWST directly measured a water plume erupting from Enceladus’ surface and measured the resulting torus of water vapor tracing its orbit around Saturn \cite{villanueva_jwst_2023}. Additionally, while definitive proof is elusive, there have been multiple tentative observations of water plumes erupting from Europa (e.g. \cite{paganini_measurement_2020}). However, these searches indicate that plume activity on Europa is either regionally limited or sporadic – making any positive detection by a low-overhead twilight monitoring program an enormous contribution to our understanding of Europa.

SCALES may be sensitive to large geysers on Enceladus, given its similar wavelength coverage, spectral resolution, and sensitivity to the strongest molecular features, but offers enhancements compared to JWST in spatial resolution and ability to do high-cadence monitoring through twilight observations. Finer spatial resolution compared to JWST/NIRSpec by a factor of $\sim$5 allows SCALES to do more detailed characterization of the variation of plume shape and composition over time. Multiple molecular features of compounds of astrobiological interest – organics, cyanide compounts – could be characterized by SCALES if present in sufficient concentrations. Cadenced monitoring with the IFU could trace evolution in volume, composition, and spatial distribution of both the plume and the larger water torus. Twilight monitoring of Enceladus may be able to characterize the spatial evolution of the plumes and torus, pinpoint eruption times, and map ejection sites across the globe. Better constraints on plume parameters and their time variability could reveal key insights into interior dynamics of Enceladus' hidden seas and advance the community priority of better understanding ocean worlds.


\subsection{Exoplanet snapshot survey}
\label{subsec:exoplanet}


The field of exoplanet direct imaging is pushing towards deeper sensitivities in a race to detect the faintest, smallest, coldest, closest-separation exoplanets. This push results in a proliferation of deep surveys utilizing long exposures and clever observational and post-processing techniques to push detections as far as they can go. However, we find that a shallow exoplanet direct imaging survey, conducted with the SCALES instrument during twilight time, is sufficiently sensitive to wide-separation giant planets and could place meaningful constraints on the occurrence rates of these objects.

Throughout the year, the period between 12$^{\circ}$ and 3$^{\circ}$ twilight (at which point sunlight is refracted above the horizon) averages to a duration of $\sim$40 minutes. Even assuming some operational overhead, this means that science programs may be able to achieve maximum exposure times of $\sim$30 minutes during twilight observing. 

We find that 30-minute exposures are sufficient for an exoplanet direct imaging survey as a SCALES twilight science program. While the raw contrasts of exoplanet direct imaging are prohibitively shallow for all but the deepest exposures (and thus too long for a twilight program), post-processing techniques can be utilized to make twilight direct imaging a possibility. We assume that twilight observations will not contain enough parallactic angle evolution to make angular differential imaging (ADI) possible; however, reference star differential imaging (RDI) may be possible for a long-running direct imaging survey, where a library of reference stars is built over time from the science exposures themselves. This will be explored during SCALES commissioning. 

 We expect SCALES to achieve similar sensitivities and contrast as NIRC2, which achieves contrasts of $C_F \sim 10 ^{-3}$ ($C_m \sim 7.5$ magnitudes) with $\lesssim$30-minute exposures assuming RDI (\cite{xuan_characterizing_2018}, see Figure 11). Because SCALES contains an IFU in addition to an imaging channel, SCALES IFU observations can leverage spectral differential imaging (SDI), which is expected to augment achievable contrast by $\sim 1$ magnitude, resulting in an estimated twilight contrast of $C_F \sim 4 \times 10 ^{-4}$ ($C_m \sim 8.5$ magnitudes) \cite{sallum_slicer_2023}. This estimate is for close separations of $\sim$0.25-1\rq\rq, where we expect to find the majority of giant companions based on population studies. However, contrast improves at larger separations, and we take $C_m \sim 8.5$ as a lower bound of SCALES twilight contrast performance.

\begin{figure}
    \centering
    \includegraphics[width=\textwidth]{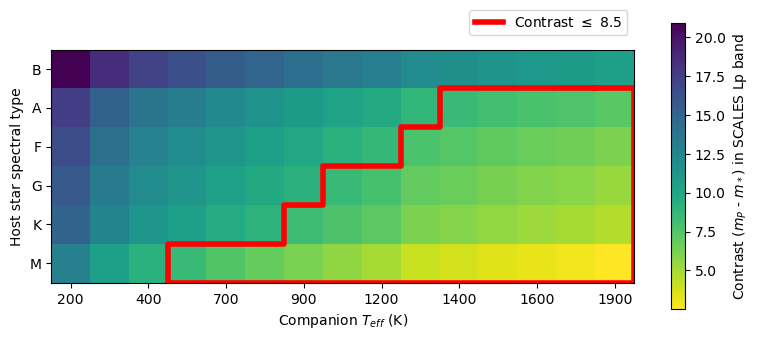}
    \caption{Host-companion contrast calculated for a range of stellar spectral types and companion effective temperatures. We derive an expected contrast floor that SCALES can achieve during twilight (30-minute exposures, with a vortex coronagraph, using RDI) at separations down to $\sim$0.25\rq\rq.}
    \label{fig:contrast}
\end{figure}

We find that with these contrast and sensitivity constraints, we expect SCALES to be sensitive to a large range of substellar companions. Figure \ref{fig:contrast} shows host-companion contrasts for a range of stellar spectral types (y-axis, calculated from Phoenix stellar models \cite{allard_models_2012}) and sub-stellar companion temperatures (x-axis, calculated from Sonora Bobcat model spectra \cite{marley_sonora_2021}). We find that a significant amount of parameter space is accessible to a SCALES twilight survey. Further work is needed to explore the mass sensitivities implied by these host and companion effective temperatures.

Current population studies of giant planets is converging on an occurrence rate of $\sim 1\%$ across stellar host demographics and semimajor axis \cite{bowler_occurrence_2018, nielsen_gemini_2019, fernandes_hints_2019, wittenmyer_cool_2020, fulton_california_2021}. 

To estimate the science yield of a SCALES twilight exoplanet survey, we estimate the number of triggers we expect per year. Table \ref{tab:triggers-per-year} shows that, with the current opportunistic twilight observing framework and assuming a single science program, SCALES could observe 112$\pm$64 potential hosts per year. This survey, operating over a 10-year lifespan, could expect to observe $\sim$1000 potential hosts for wide-separation giant planets; taking a 1\% occurrence rate from the literature, we could expect to detect $\sim$11$\pm$6 wide-separation giant companions. More work is needed to constrain the expected performance of a SCALES twilight exoplanet survey, and the observing tools to enable program operations will be developed during commissioning in early 2026.

\renewcommand{\arraystretch}{1.2}
\begin{table}[H]
\begin{tabular*}{\columnwidth}{@{\extracolsep{\fill}}l ccc}
\toprule
\toprule
     & \textbf{Both} & \textbf{K1} & \textbf{K2} \\
  \cmidrule(lr){2-4}
  1. The Twilight Zone & 24$\pm$5 & 4$\pm$2  & 20$\pm$4 \\
  2. The Twilight Zone, re-appropriating some early shutdowns     & 53$\pm$7 & 16$\pm$4  & 38$\pm$6 \\
  3. Facilitized twilight observing program & 111$\pm$64 & 112$\pm$64  & 223$\pm$127 \\ 
\bottomrule
\bottomrule
\end{tabular*}{
\caption{Number of nights per year utilized by the three twilight program frameworks considered in this paper. This table is functionally identical to Table \ref{tab:time-results}, but phrased in terms of number of twilight program triggers per year (assuming a single twilight observation per morning twilight) instead of accumulated time spent collecting science photons.}
\label{tab:triggers-per-year}
}
\end{table}

\section{Roadmap to a facilitized Keck twilight observing program}
\label{sec:recommendations}

For a facilitized twilight observing program that takes advantage of the full 151$\pm$2 hours per year of additional Keck observing  (as estimated in Section \ref{sec:time}), there are a number of programmatic and technical tasks that must be undertaken by Keck Observatory staff, Keck instrument teams, and the Keck user community. The observatory and the user community must jointly agree upon and work together on the implementation of these changes for overall programmatic success and an increase in science yield.

\textbf{(i) Keck Observatory could seek to move all non-science activities out of science-capable twilight time.} Twilight time is an underutilized resource that offers a unique observing mode. However, because it is non-optimal observing time for optical instruments, it is often appropriated not for science but for instrument calibrations or facility operations. Proper instrument calibration is extraordinarily important, but calibration procedures are non-standardized across the user community, which risks wasting science-capable time on an unnecessary volume of instrument calibrations. \textbf{Keck Observatory could conduct a study of facility instrument calibrations.} This includes studying which calibrations can be taken during non-science-capable time or entirely off-sky (e.g. all SCALES calibrations are internal to the instrument and are automatically executed during the day); and studying the optimal number and type of calibrations to ensure excellent data quality while preserving science photons (e.g. perhaps ). In addition to optimizing and streamlining instrument calibrations, \textbf{Keck Observatory could conduct a study of which facility activities can be moved out of science-capable twilight time.} 

\textbf{(ii) The Keck user community should consider and propose for a complementary set of twilight science programs to maximally utilize this unique observing mode.} As discussed in Section \ref{sec:science}, while Solar System monitoring is an excellent fit for twilight observing given the time-domain nature, it leaves more than half of annual twilight time unused given the seasonal inaccessibility of Solar System targets. This paper proposes an exoplanet direct imaging survey campaign (see Section \ref{subsec:exoplanet}), but the community should study and propose for a range of science programs that take full advantage of the time and unique capabilities made available by this observing mode.

\textbf{(iii) The Keck user community should collaborate with Keck Observatory staff to optimize AO controls for operations under a brightening twilight sky.} This work is already under way. While the infrared sky remains at a mostly stable brightness throughout twilight, the R-band sky experiences a change in brightness of several magnitudes between dark time and sunrise. The Keck AO Shack-Hartmann wavefront sensor (SHWFS) operates in R-band, and struggles to maintain closed loop control on fainter guide stars as the sun approaches the horizon. It is worth noting that the incoming infrared (H-band) pyramid wavefront sensor will not face the same challenges, given the comparative stability of the twilight sky in H-band compared to R-band \cite{wizinowich_adaptive_2024}.

Currently, the SHWFS takes one measurement of the sky background and subtracts this background from each read of the SHWFS. Re-taking this background measurement can take up to $\lesssim$2 minutes, since it requires opening the loops, slewing to an offset target, taking the background measurement, returning to the original target, and re-closing loops. It is infeasible to repeatedly retake these background measurements during the time-constrained observing period of twilight. 

 However, a lightweight \lq\lq twilight mode\rq\rq  background scaling routine can extend Keck AO performance to fainter guide stars during the brighter twilight regimes by adjusting the SHWFS background image in accordance with the brightening R-band sky. We are adapting the open-source ESO SkyCalc tool \cite{noll_atmospheric_2012, jones_advanced_2013} to model sky brightness in R-band from dark time through 3$^{\circ}$ twilight (at which point Earth's atmosphere begins to refract sunlight over the horizon). This sky brightness model, which includes an approximation for twilight brightening, has been verified by the LSST instrument team for Cerro Pachon \cite{yoachim_optical_2016}, and a verification of this tool for Maunakea is in progress.  We have begun conducting on-sky testing to better parameterize current Keck AO performance during twilight, and this scaling tool will be described in future work.


\appendix    

\acknowledgments 
 
This work was enabled by the Keck Visiting Scholars program and by the NSF Graduate Research Fellowship Program (NSF GRFP). This work benefited from the 2025 Exoplanet Summer Program in the Other Worlds Laboratory (OWL) at the University of California, Santa Cruz, a program funded by the Heising-Simons Foundation and NASA. We would like to thank everyone at Keck Observatory who generously lent their time and expertise, including but not limited to Jim Thorne, Jacques Delorme, Max Brodheim, Tyler Coda, Rosalie McGurk, Sherry Yeh, Shui Kwok.

\printbibliography

\end{document}